# Magnetic field filtering of the hinge supercurrent in unconventional metal NiTe$_2$-based Josephson junctions

Tian Le[1#], Ruihan Zhang[1,2#], Changcun Li[3], Ruiyang Jiang[1,2], Haohao Sheng[1,2], Linfeng Tu[1,4], Xuewei Cao[4], Zhaozheng Lyu[1,5], Jie Shen[1,6], Guangtong Liu[1,5,6], Fucai Liu[3,7*], Zhijun Wang[1,2*], Li Lu[1,2,5,6*], Fanming Qu[1,2,5,6*]

[1] Beijing National Laboratory for Condensed Matter Physics, Institute of Physics, Chinese Academy of Sciences, Beijing 100190, China

[2] School of Physical Sciences, University of Chinese Academy of Sciences, Beijing 100049, China

[3] School of Optoelectronic Science and Engineering, University of Electronic Science and Technology of China, Chengdu 611731, China

[4] School of Physics, Nankai University, Tianjin 300071, China

[5] Hefei National Laboratory, Hefei 230088, China

[6] Songshan Lake Materials Laboratory, Dongguan, Guangdong 523808, China

[7] Yangtze Delta Region Institute (Huzhou), University of Electronic Science and Technology of China, Huzhou 313009, China

[#] These authors contributed equally to this work.

[*] Emails: fucailiu@uestc.edu.cn; wzj@iphy.ac.cn; lilu@iphy.ac.cn; fanmingqu@iphy.ac.cn

**Topological materials with boundary (surface/edge/hinge) states have attracted tremendous research interest. Besides, unconventional (obstructed atomic) materials have recently drawn lots of attention owing to their obstructed boundary states. Experimentally, Josephson junctions (JJs) constructed on materials with boundary states produce the peculiar boundary supercurrent, which was utilized as a powerful diagnostic approach. Here, we report the observations of conspicuous hinge supercurrent in NiTe$_2$-based JJs. Particularly, applying an in-plane magnetic field along the Josephson current could rapidly suppress the bulk supercurrent and retain the nearly pure hinge supercurrent, namely the magnetic**

**field filtering of supercurrent. Further systematic comparative analysis and theoretical calculations demonstrate the existence of unconventional nature and obstructed hinge states in NiTe₂. Our results revealed the unique hinge states in unconventional metal NiTe₂, and demonstrated in-plane magnetic field as an efficient method to filter out the futile bulk contributions and thereby to highlight the hinge states hidden in topological/unconventional materials.**

Bulk band topology permeates in three- and two-dimensional (3D and 2D) condensed matter, e.g., topological insulators and topological semimetals, and gives rise to gapless surface/edge states, which is a well-known bulk-boundary correspondence protected by topological invariants[1-3]. In recent years, the concept has been extended to $d$-dimensional higher-order topological systems with ($d$-$m$)-dimensional ($m≥2$) metallic hinge or corner states[4-10]. Josephson junctions (JJs) have served as a powerful tool to reveal boundary states in topological materials, where supercurrent distributions modulated by the boundary supercurrent could be discriminated by measuring the interference pattern of the critical supercurrent in a magnetic field[11-21].

On the other hand, a new category of unconventional materials has been proposed with the obstructed atomic nature, where the electrons are located away from the nuclei in crystals[22-29]. As a result of the mismatch between average electronic centers and atomic positions, the obstructed states emerge on the boundary, whose bulk band gaps could be much larger than those of topological materials. In general, both the topological and obstructed boundary states can be used for constructing topological superconductivity and Majorana zero modes with the assistance of superconducting proximity effect (SPE)[30-33]. However, in principle it is a challenge to distinguish the boundary states hidden in a semimetal/metal from bulk states, because both of them are metallic. But for their JJs, the supercurrent on bulk states is expected to suffer larger decoherence and dephasing effects than the boundary states, and therefore, the JJs would exhibit a particular behaviour based on the boundary supercurrent channels[12-16].

In this work, we report the observations of the hinge supercurrent in unconventional metal NiTe$_2$-based JJs. Particularly, an in-plane magnetic field (only few tens of millitesla) applied parallel to the Josephson current could filter out the bulk supercurrent and retain the robust hinge supercurrent. Based on a further comparison with a JJ which did not include the hinges of the sample, the effect of an in-plane magnetic field perpendicular/parallel to the Josephson current, and theoretical calculations, these observations could be attributed to obstructed hinge states in the unconventional metal NiTe$_2$. Especially, the magnetic field filtering of the supercurrent functions as a compelling route to acquire the nearly pure boundary supercurrent in topological/unconventional materials-based JJs.

NiTe$_2$ crystallizes in the CdI$_2$-type trigonal structure with a $P\bar{3}m1$ space group (number 164), as schematically illustrated by the left inset of Fig. 1a. The NiTe$_2$ layers individually stack along the *c*-axis ($C_3$ rotation axis) via van der Waals (vdW) force. It was reported as a type-II Dirac material by *ab initio* calculations and angle-resolved photoemission spectroscopy measurements[34,35]. Exfoliated NiTe$_2$ nanoplates with a thickness more than 30 nm were used in this work. Figure 1a shows the temperature dependence of the resistance ($R$) for a NiTe$_2$ nanoplate from room temperature to 1.55 K with a typical metallic behaviour. The magnetic field dependence of $R$ suggests a nearly non-saturating linear or sublinear magnetoresistance as shown by the right inset of Fig. 1a, which is similar to the bulk materials[34,36].

We fabricated JJs on NiTe$_2$ nanoplates with superconducting electrodes NbTiN as shown in Fig. 1b. JJs of this type have been successfully implemented on many topological materials to explore the boundary states[13,14,37-40]. The current-voltage (*I-V*) characteristics of device D1 is shown in Fig. 1c, indicating a Josephson critical supercurrent ($I_c$) of ~1 μA. When a magnetic field is applied perpendicular to the junction ($B_z$), the superconducting interference pattern (SIP) could be obtained as illustrated in Fig. 2a. The SIP is characterized by the periodic oscillations of $I_c$ as marked by the whitish envelope which separates the superconducting and normal states. We note that the $I_c$ decays very slowly with increasing $|B_z|$, which is in stark contrast to the standard one-slit Fraunhofer-like pattern with the form $|sin(\pi\Phi/\Phi_0)/(\pi\Phi/\Phi_0)|$ in conventional JJs, denoted by the red line in Fig. 2a, where $\Phi = L_{eff}WB_z$ is the magnetic flux, $L_{eff}$ and $W$ are the effective length and width of the junction, respectively, and $\Phi_0 = h/2e$ is the flux quantum ($h$ is the Planck constant, $e$ is the elementary charge)[11]. A similar phenomenon has been reported on various materials which was attributed to the large boundary (edge or hinge)–supercurrent density in the JJs[13-21]. As for JJs, the supercurrent density $J_s$ as a function of position $y$, $J_s(y)$, can be extracted from the $B_z$ dependence of $I_c$, $I_c(B_z)$, through the Fourier transform (Dynes-Fulton approach)[41]. Figure 2b depicts the supercurrent density profile $J_s(y)$ extracted from the $I_c(B_z)$ curves, retrieved from Fig. 2a accordingly (see Supplementary Section I). The center of the junction corresponds to the position $y = 0$. Note that large supercurrent densities appear around $y = \pm 0.7$ μm, which locate at the hinges or side surfaces of the sample and give rise to the boundary supercurrent. Therefore, the SIP on D1 is constituted by the bulk and hinge/side-surface supercurrent. In the following, boundary refers to the hinges or side surfaces of the NiTe2 nanoplate.

We next investigate the effect of an in-plane magnetic field on the SIP. Figures 2c and 2d show the SIPs under $B_x = 0.1$ T and 0.2 T, respectively. Here, $B_x$ denotes the in-plane magnetic field parallel to the Josephson current, while $B_y$ is perpendicular to the Josephson current. The width of the central lobe decreases when increasing $B_x$, as indicated by the red dashed line. Notably, the relative height of the central lobe to the side lobes is strongly suppressed at $B_x = 0.2$ T, indicating a boundary-dominant supercurrent. As sketched in Fig. 2e, when the bulk supercurrent is dominant, it presents a standard Fraunhofer-like pattern with the central lobe possessing a width of $2\Phi_0$ and the side lobe of $\Phi_0$. In particular, the height of the lobes shows a global $1/|B_z|$ fast decay (top row). On the contrary, when the supercurrent flows only along the two hinges/side surfaces (bottom row), the single JJ imitates a superconducting quantum interference device (SQUID) with a two-slit SIP which follows the form $|cos(\pi\Phi/\Phi_0)|$. In this case, both the central and side lobes have a uniform width of $\Phi_0$ and a weak global decay. If considering the admixture of bulk and boundary supercurrent, the SIP has a finite weight of SQUID signal (middle row), which exactly corresponds to the SIP on our NiTe$_2$-based JJ D1 at $B_x = 0$ T. Accordingly, as shown in Fig. 2d, with the increase of $B_x$ the contribution from the bulk supercurrent decreases significantly and finally a SQUID-like pattern emerges with the boundary-dominant supercurrent. Therefore, the in-plane magnetic field $B_x$ could filter out the bulk supercurrent, i.e., the magnetic field filtering of supercurrent is observed in our experiments.

Such filter effect is crucial for inspecting the contribution from the boundary supercurrent in JJs, even if the weight of the boundary supercurrent is not large enough. Figures 3a and 3c depict the SIPs for devices D2-1 and D3-1 with different weights of the boundary supercurrent without in-plane magnetic fields. Here, D2-1 displays the deviation from the standard Fraunhofer-like pattern (red line) mainly on the first and second side lobes, while D3-1 shows a very little deviation. The supercurrent density profiles of Figs. 3a and 3b shown in the right insets indicate the small weight of the boundary supercurrent. However, an in-plane magnetic field $B_x = 0.04$ T entirely kills the bulk supercurrent and yields SQUID-like patterns as shown in Figs. 3b and 3c ($B_x$ for killing the bulk supercurrent is device dependent; see Supplementary Section II). In the same way, the supercurrent density profiles in the insets of Figs. 3b and 3c illustrate the dominance of the boundary supercurrent.

We further compared the effect between $B_x$ and $B_y$ on D4-1 and realized that $B_y$ has a negligible filter effect on the bulk supercurrent. Without the in-plane magnetic field, D4-1 exhibits a SIP with a bulk-dominant supercurrent, as plotted in Fig. 4a. Applying $B_x = 0.04$ T is successful in presenting the SQUID-like pattern as shown in Fig. 4b, as expected. However, the SQUID-like pattern is always absent for $B_y$ = 0.04 T, 0.06 T and 0.2 T, as shown in Figs. 4c, 4d and 4e, respectively (The different critical magnetic field of the bulk supercurrent between $B_x$ and $B_y$ is discussed in Supplementary Section III).

Regarding to the origin of the observed boundary supercurrent, it is commonly attributed to the proximity-induced superconductivity on hinge/side-surface channels[13-21]. However, the bending of the magnetic field lines around the edges of the electrodes was also proposed[42]. In order to further clarify the essential role of the sample hinges/side surfaces, we fabricated a JJ whose junction region did not include the hinges/side surfaces of the sample, as shown by the left inset of Fig. 5a (device D2-2; the upper left junction of D2 shown in the inset of Fig. 3a). The SIP only presents a central lobe (Fig. 5a), which corresponds to the supercurrent density decaying from the center to the edges of the JJ region, as depicted in the right inset of Fig. 5a. Consistently, the SIP does not show any SQUID-like signal even if applying a magnetic field $B_x$ = 0.05 T, as displayed in Fig. 5b (The width of central lobe is smaller than Fig. 5a, primarily due to the large suppression of supercurrent at $B_x$ = 0.05 T, which can also be seen in Fig. 4e at $B_y$ = 0.2 T). It would be a critical evidence to pin down the role of the sample hinges/side surfaces for the boundary supercurrent in our JJs.

Next, we will demonstrate that the boundary supercurrent comes from the sample hinges. If there is a SPE on the side surfaces, the damping of the critical boundary supercurrent (boundary-$I_c$) under $B_y$ follows the Fraunhofer-like curve. Note that the $1/|B_z|$ global decay, which is a manifestation of the cancellation of the positive and negative supercurrent, helps to supress the contribution of the bulk supercurrent for the side lobes[11]. Considering a nanoplate thickness of ~30 nm, and a separation of the electrodes ~300 nm, boundary-$I_c(B_y)$ for the side lobes (n) could be calculated from Fig. 2a for device D1 assuming a side-surface supercurrent, as shown by the pink triangles in Fig. 5c (detailed analysis is shown in Supplementary Section IV). However,

boundary-$I_c(B_y = 0.2$ T) (blue squares) extracted from side lobes of Fig. 2d are much larger than the calculated values (pink triangles), as shown in Fig. 5c, which is in contrast with a side-surface supercurrent. Furthermore, boundary-$I_c(B_y = 0.2$ T) are comparable to the boundary-$I_c$ at $B_x = 0.2$ T (red circles), which does not support the side-surface scenario, either, since a larger suppression of the supercurrent caused by the Fraunhofer-like decay and the probable orbital effect is expected for $B_y$ if assuming a side-surface supercurrent. Therefore, it points to the hinge supercurrent.

We next investigate the origin of the hinge states. Comparing with the hinge supercurrent originating from the higher-order topology in $Cd_3As_2$ and $WTe_2$-based JJs[13-16], it is intriguing to scrutinize the topological hinge states in $NiTe_2$. However, the type-II Dirac point in $NiTe_2$ is embedded in the bulk bands, and there is no clue yet that it could present topological hinge states.

Instead, our detailed calculations show that $NiTe_2$ has the unconventional nature of charge mismatch, which gives rise to the obstructed hinge states. Meanwhile, the locked spin of the hinge states could explain the observed magnetic field filtering effect of the hinge supercurrent (as shown later). We calculated the $NiTe_2$ rod and obtained the projected spectrum on the hinge atoms shown in the inset of Fig. 6f. Since it is a vdW layered compound, we tried to investigate the monolayer for convenience. We slightly enlarged the interlayer Te-Te distance (only modifying Te-$p_z$ dispersion) and computed the orbital-resolved band structures (Figs. 6a, b and c) and Wannier charge centers (Fig. 6d). The results show that the Te-$p_{x, }p_y$ and Ni-$d$ orbitals have a strong hybridization. The $Te^{2-}$ valence state usually means that the Te-$p$ orbitals are fully occupied. Surprisingly, there is a large weight of Te-$p_x/p_y$ orbitals in the conduction bands, which contradicts with the $Te^{2-}$ valence state. On the other hand, using $z$-directed 1D Wilson loop technique, Wannier charge centers (WCC) are obtained (Fig. 6d), and the two average charge centers in the red box are quite away from the Te atoms (the dashed lines), indicating the unconventional nature of $NiTe_2$ monolayer. Then, when we performed the calculation in an open boundary condition, the obstructed states were obtained on the edge (Fig. 6e) (red and blue bands indicate the different spin channels due to spin-orbit interactions). To investigate the side surface and hinge states of the bulk NiTe$_2$, we have performed a rod calculation with open boundary conditions in both *b* and *c* directions. The results in Fig.6f show the hinge states clearly (projected onto the hinge atoms in the red box), while the side-surface states are much weaker than the hinge states due to the interlayer hybridization (see Fig.S4 in the Supplementary Section VI). Thus, we demonstrated that the hinge states are regarded as the remanence of the obstructed states of the unconventional metal NiTe$_2$[26,27]. Intriguingly, such unconventional materials were also found to be suitable to construct Josephson diode[25,26,43].

Due to the existence of time reversal symmetry and mirror symmetry (m$_x$), the electron spins of the hinge states are locked to be in the plane perpendicular to the hinges (along the *x* direction), which could explain the filter effect of the supercurrent under $B_x$. The 1D hinge channel undergoes less scattering than the bulk[13] due to the locked spin, and thereby produces a more robust supercurrent under the in-plane magnetic field $B_x$. However, $B_y$ could act the Lorentz force on the hinge channel and weaken the hinge supercurrent. Therefore, the magnetic field filtering of supercurrent is absent for $B_y$.

In conclusion, we uncovered the hinge supercurrent in NiTe$_2$-based JJs. Our observations combined with the theoretical calculations revealed the unconventional nature and hinge states in NiTe$_2$. In particular, we demonstrated the in-plane magnetic field filtering as a route of vital importance to eliminate unserviceable contributions from bulk states in topological/unconventional materials with hinge states.

# Methods

**Crystal growth and device fabrication.** High-quality NiTe$_2$ single crystals were synthesized using the Te flux method. High purity Ni powder (99.9%) and Te ingots (99.99%) with a ratio of 1:10 were sealed in an evacuated quartz tube. Then, the sealed quartz tube was placed in a furnace, and heated to 1000 °C over 10 hours. After staying at a constant temperature of 1000 °C for 10 hours, the tube was cooled to 600 °C in a rate of 3 °C/h and kept at 600 °C for 3 days to improve the quality of the single crystals. Finally, single crystals of NiTe$_2$ were obtained by removing the remaining Te flux at 600 °C. NiTe$_2$ nanoplates were exfoliated from the single crystals onto a Si/SiO$_2$ wafer in air atmosphere. Ti/NbTiN (5 nm/100 nm) electrodes were deposited via magnetron sputtering with a preceding soft plasma cleaning, using standard electron-beam lithography techniques.

**Transport measurements.** Measurements of the temperature and magnetic field dependence of the resistance of NiTe$_2$ nanoplates were carried out with a four-probe configuration in an Oxford TeslatronPT system equipped with a 14 T superconducting magnet. The measurements of Josephson junctions were carried out with a quasi-four-probe configuration in cryofree Oxford Triton dilution refrigerators equipped with a 3-axis vector magnet. A source meter (Keithley 2612B) was used to apply a d.c. bias current $I_{dc}$. A lock-in amplifier (LI5640, NF Corporation) was used to apply a small a.c. excitation current $I_{ac}$ and obtain the differential resistance d$V$/d$I$ = $V_{ac}$/$I_{ac}$.

**Theoretical calculation.** We performed the first-principles calculations within the framework of the density functional theory (DFT) using projector augmented wave (PAW) method[44,45] implemented in the Vienna *ab initio* simulation package (VASP)[46,47]. The generalized gradient approximation (GGA) of Perdew-Burke-Ernzerhof (PBE) type[48] was employed for the exchange-correlation potential. The kinetic energy cutoff for plane wave expansion was set to 400 eV. The thickness of the vacuum layers along *b* and *c* directions for the NiTe$_2$ bulk and monolayer with open boundary conditions were set to > 20 Å. The Brillouin zone was sampled by Γ-centered Monkhorst-Pack method in the self-consistent process, with a 9×9×6 $k$-mesh for NiTe$_2$ bulk and a 10×1×1 $k$-mesh for the NiTe$_2$ bulk and monolayer with open boundary conditions.

## Data availability

The data that support the findings of this study are available from the corresponding authors on reasonable request.

## Acknowledgements

We would like to thank Yi Zhou for fruitful discussions. This work was supported by the National Key Research and Development Program of China (2022YFA1403400, 2022YFA1403800, 2021YFE0194200, 2020YFA0309200, 2017YFA0304700); by the National Natural Science Foundation of China (12074417, 92065203, 11974395, 12188101, 12104489, and 11527806); by the Strategic Priority Research Program of Chinese Academy of Sciences (XDB28000000, and XDB33000000); by the Synergetic Extreme Condition User Facility sponsored by the National Development and Reform Commission; and by the Innovation Program for Quantum Science and Technology (2021ZD0302600); and by the Center for Materials Genome.

## Author contributions

F.Q. supervised the project. C.L. synthesized the crystals under the instruction of F.L. T.L., R.J. and L.T. fabricated the devices and performed the transport measurements, with the support from X.C., Z.L., J.S., G.L., L.L. and F.Q. T.L. and F.Q. analysed the data with the help from Z.W. and L.L. R.Z., H.S. and Z.W. did the DFT calculations and symmetry analysis. T.L., F.Q., Z.W., R.Z., and L.L. wrote the manuscript, with input from all authors.

## Competing interests

The authors declare no competing interests.

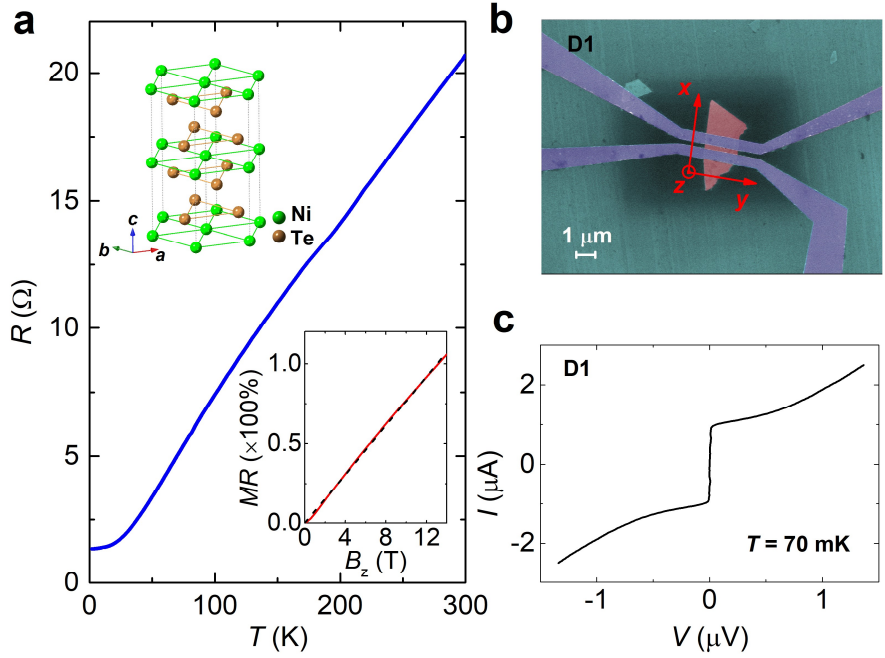

**Fig. 1. Characterization of the NiTe$_2$-based JJ. a**, Temperature dependence of resistance $R$ of an exfoliated NiTe$_2$ nanoplate. The left inset is a schematic illustration of the atomic structure of NiTe$_2$ crystal. The right inset is magnetic field dependence of $R$ at 1.55 K. The black dashed line represents a linear fit to the data. **b**, False-colour scanning electron microscopic image of a typical NiTe$_2$-based Josephson junction D1, where the purple colour represents superconducting NbTiN electrodes with a width $t \sim$ 500 nm. The separation $L$ between electrodes is ~300 nm. The width $W$ of the NiTe$_2$ nanoplate (red colour) between the two electrodes is ~1.5 μm. **c**, The $I$-$V$ characteristic curve showing the Josephson supercurrent at 70 mK.

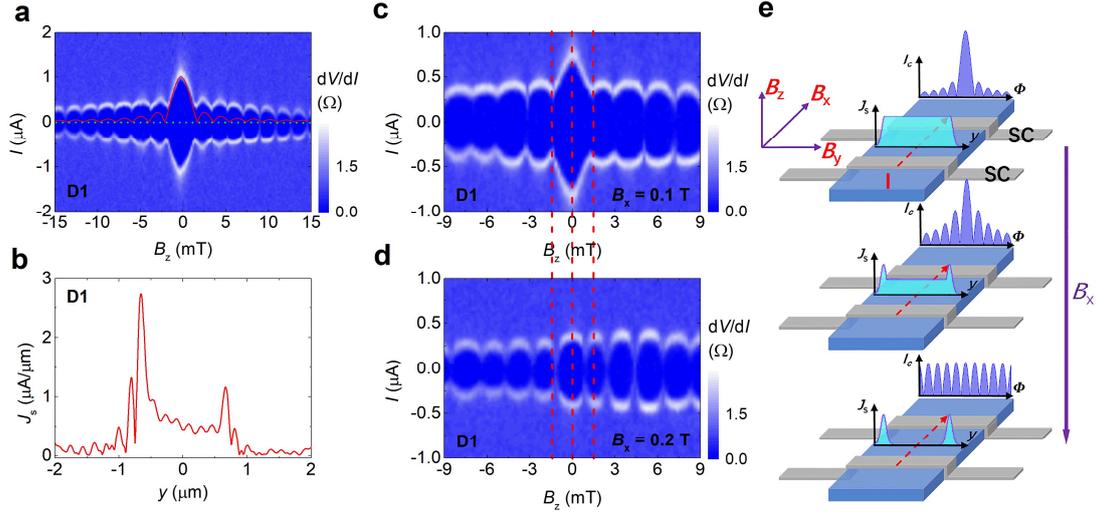

**Fig. 2. In-plane magnetic field $B_x$ tuning for the SIP. a**, SIP for D1 at 70 mK. The red line is a standard Fraunhofer-like curve. **b**, Supercurrent density profile $J_s(y)$ for D1 after the Fourier transform of $I_c(B_z)$ in **a**. **c**, **d**, SIP for D1 at $B_x$ = 0.1 T and 0.2 T, respectively. **e**, Schematic illustration of $B_x$-tunable supercurrent density distribution and SIP.

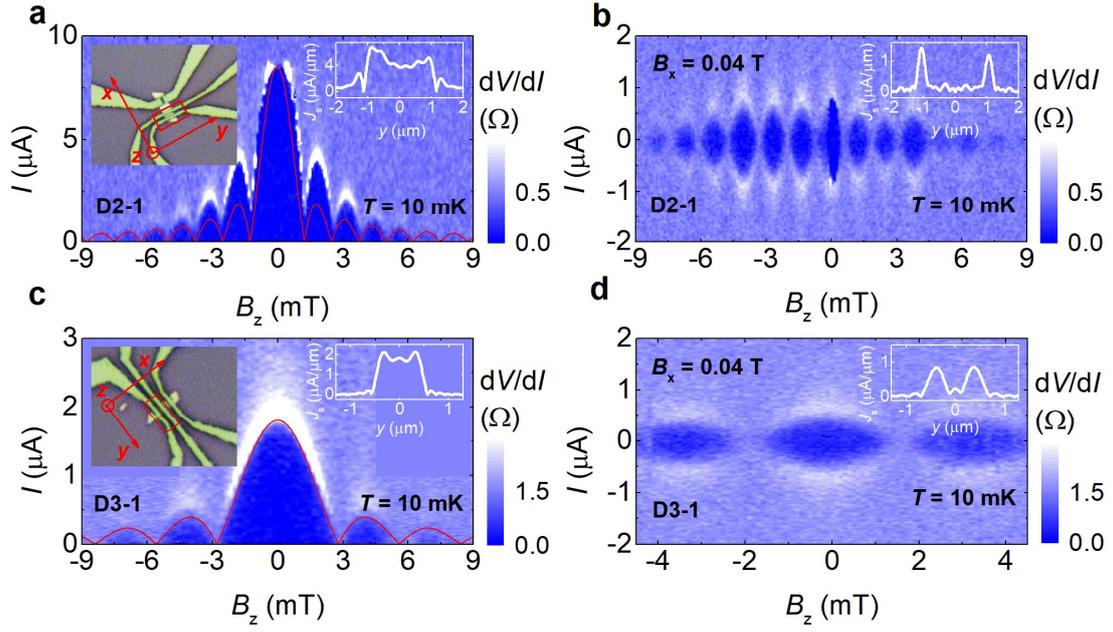

**Fig. 3. In-plane magnetic field $B_x$ filtered boundary supercurrent on junctions D2-1 and D3-1. a, c**, SIP for D2-1 and D3-1 at 10 mK without the in-plane magnetic field, respectively. The red line represents the standard Fraunhofer-like curve. The left insets are optical images for D2-1 and D3-1, indicated by red frames, and with the sample width of 1.9 µm and 1.1 µm, respectively. The right insets depict supercurrent density profiles $J_s(y)$. **b, d**, SIP for D2-1 and D3-1 at 10 mK under $B_x = 0.04$ T, respectively. The insets are corresponding supercurrent density profiles $J_s(y)$.

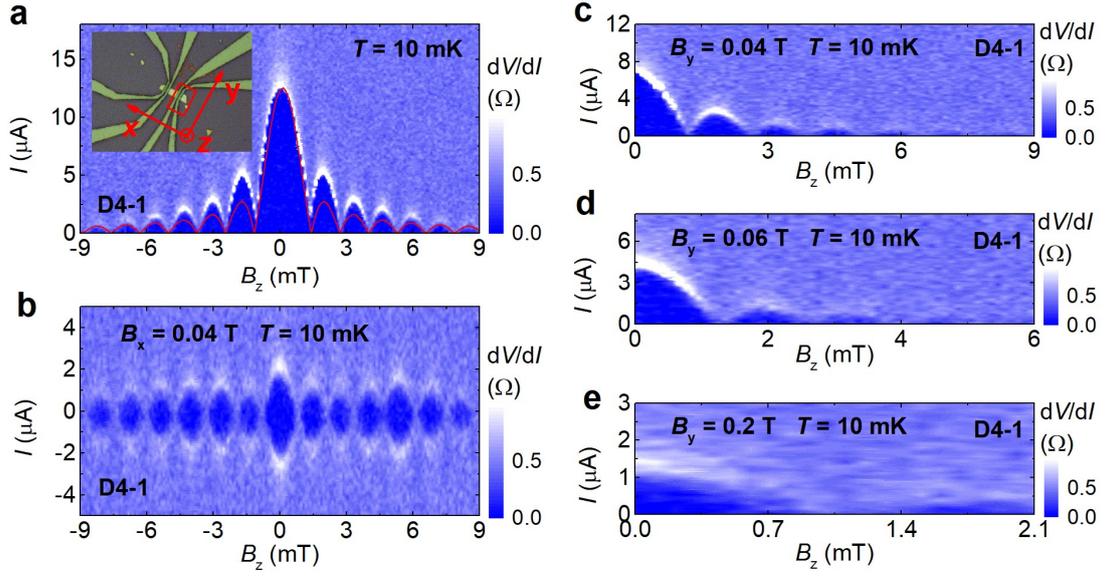

**Fig. 4. Comparison of the effect of $B_x$ and $B_y$ on SIP for D4-1. a**, SIP for D4-1 at 10 mK without the in-plane magnetic field. The red line represents the standard Fraunhofer-like curve. The inset displays the optical image for D4-1, indicated by the red frame, and with the sample width of 2.1 μm. **b**, SIP for D4-1 at 10 mK under $B_x$ = 0.04 T. **c, d, e**, SIP for D4-1 at 10 mK under $B_y$ = 0.04 T, 0.06 T and 0.2 T, respectively.

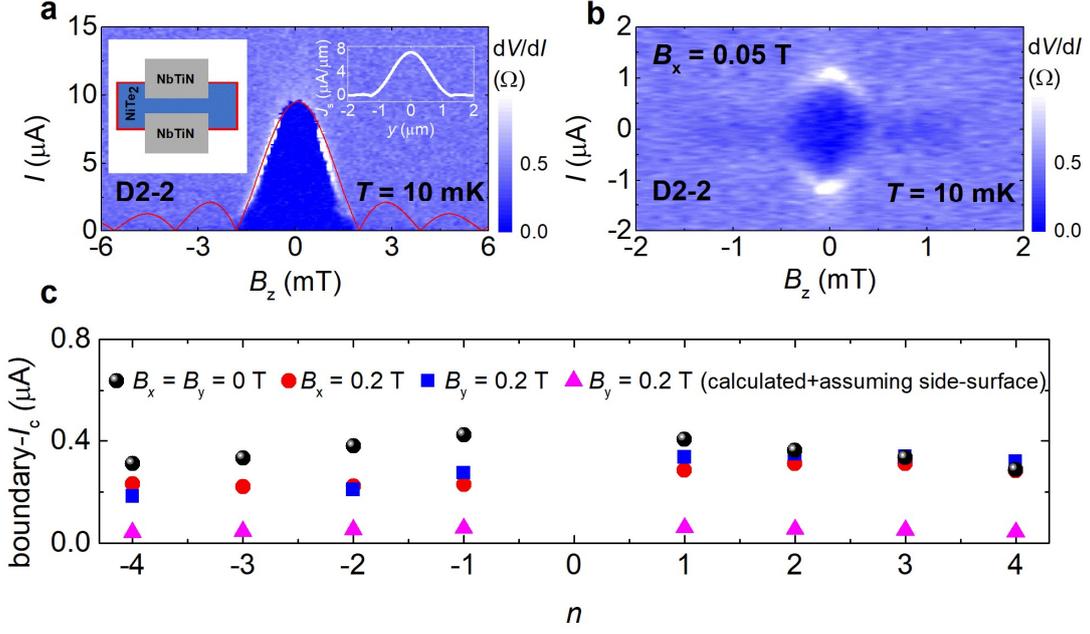

**Fig. 5. Evidence for hinge supercurrent. a**, SIP for D2-2 at 10 mK without the in-plane magnetic field. The red line represents the standard Fraunhofer-like curve. The left inset is the schematic illustration of D2-2. The right inset shows the supercurrent density profile $J_s(y)$. **b**, SIP for D2-2 at 10 mK under $B_x = 0.05$ T. **c**, The black balls denote the extracted boundary-$I_c$ from each center of the side lobes in Fig. 2a with $B_x = B_y = 0$ T. The red cycles and blue squares denote the extracted boundary-$I_c$ from each center of the side lobes in Fig. 2d and Fig. S4 with $B_x = 0.2$ T and $B_y = 0.2$ T, respectively. The pink triangles represent the calculated boundary-$I_c$ for each center of side lobes when $B_y = 0.2$ T, if assuming the existence of side-surface supercurrent. $n$ denotes the serial number of the side lobes.

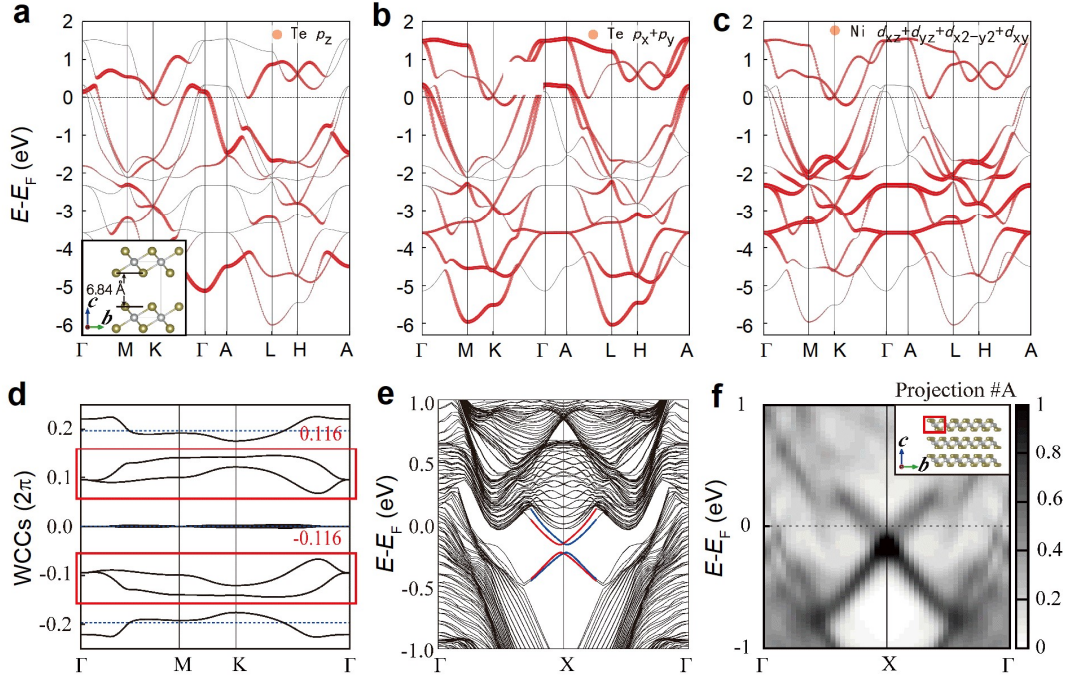

**Fig. 6. Calculation of the obstructed hinge states. a, b, c,** The orbital-resolved band structure for NiTe$_2$ (with $d_z$ = 6.84 in the inset), which shows the strong hybridization between the Te-$p_x$, $p_y$ orbitals and Ni-$d$ orbitals. **d,** The $z$-directed Wannier charge centers for the occupied nine bands. The dashed lines (0.19c) indicate the locations of the Te atoms. Two Wannier charge centers in the red box are quite away from the Te atoms, with the average of 0.116c. It indicates that the NiTe$_2$ layer is unconventional with mismatched electronic charge centers. **e,** The obtained obstructed states of the NiTe$_2$. The hinge states are highlighted in the red and blue lines. **f,** The hinge spectrum of the NiTe$_2$ bulk with open boundary conditions in both $b$ and $c$ directions. The inset shows the projected atoms on the hinge.

# Supplementary Information

## Content

**Section I. Extraction of the supercurrent density profile.**

**Section II. Magnitude of $B_x$ to kill the bulk supercurrent for different devices.**

**Section III. Anisotropy of bulk supercurrent between $B_x$ and $B_y$.**

**Section IV. Analysis of the critical side-surface supercurrent (side-surface-$I_c$) under $B_y$.**

**Section V. SIP for D1.**

**Section VI. Theoretical calculation.**

**Section I. Extraction of the supercurrent density profile.**

In this section, we will introduce the Dynes-Fulton approach for converting the superconducting interference pattern (SIP) $I_c(B_z)$ to the supercurrent density profile $J_s(y)$, taking Figs. 2a and 2b in the main text as an example.

In our configurations, the supercurrent density varies along the $y$ direction. The separation of the two electrodes is $L$ and the width of each electrode is $t$. Considering the magnetic flux focusing effect of the superconducting electrodes, the effective length of the junction is $L_{\text{eff}} = L+t$. The critical supercurrent $I_c(B_z)$ is extracted following the whitish envelope in Fig. 2a which was replotted as Fig. S1a. The red curve in Fig. S1a illustrates $I_c(B_z)$. The experimentally observed $I_c(B_z)$ is the magnitude of the integration of $J_s(y)$. Assuming the normalized magnetic field unit of $\kappa = 2\pi L_{eff} B_z/\Phi_0$, where $\Phi_0 = h/2e$ is the flux quantum ($h$ is the Planck constant and $e$ is the elementary charge), $I_c(B_z)$ can be replaced by the complex critical supercurrent function $\Im_c(\kappa)$:

$$I_c(\kappa) = |\Im_c(\kappa)| = \left|\int_{-\infty}^{\infty} J_s(x) e^{i\kappa y} dy\right|.$$

Considering an even supercurrent density $J_{even}(y)$ with a symmetric distribution, the odd part of $e^{i\kappa y}$ vanishes from the integral, and we can obtain

$$\Im_c(\kappa) = I_{even}(\kappa) = \int_{-\infty}^{\infty} J_{even}(y) \cos(\kappa y) dy.$$

$\mathfrak{I}_c(\kappa)$ in the above formula alternates between positive and negative values at each zero-crossing. Because $I_{even}(\kappa)$ usually dominates in the observed critical supercurrent except at its minima, it can be roughly obtained by multiplying $I_c(\kappa)$ with a flipping function as shown in Fig. S1b, where the sign is switched between adjacent lobes of the whitish envelope. When $I_{even}(\kappa)$ is minimal, $I_c(\kappa)$ is dominated by the odd component: $I_{odd}(\kappa) = \int_{-\infty}^{+\infty} J_{odd}(y)\sin(\kappa y)dy$. $I_{odd}(\kappa)$ can then be approximated by interpolating between the minima of $I_c(\kappa)$, and flipping sign between lobes as shown in Fig. S1c. A Fourier transform of the complex critical supercurrent function $\mathfrak{I}_c(\kappa) = I_{even}(\kappa) + iI_{odd}(\kappa)$ is:

$$J_S(y) = \left| \frac{1}{2\pi} \int_{-W/2}^{W/2} \mathfrak{I}_c(\kappa) e^{-i\kappa y} d\kappa \right|,$$

which produces the supercurrent density profile as shown in Fig. S1d (the same as Fig. 2b).

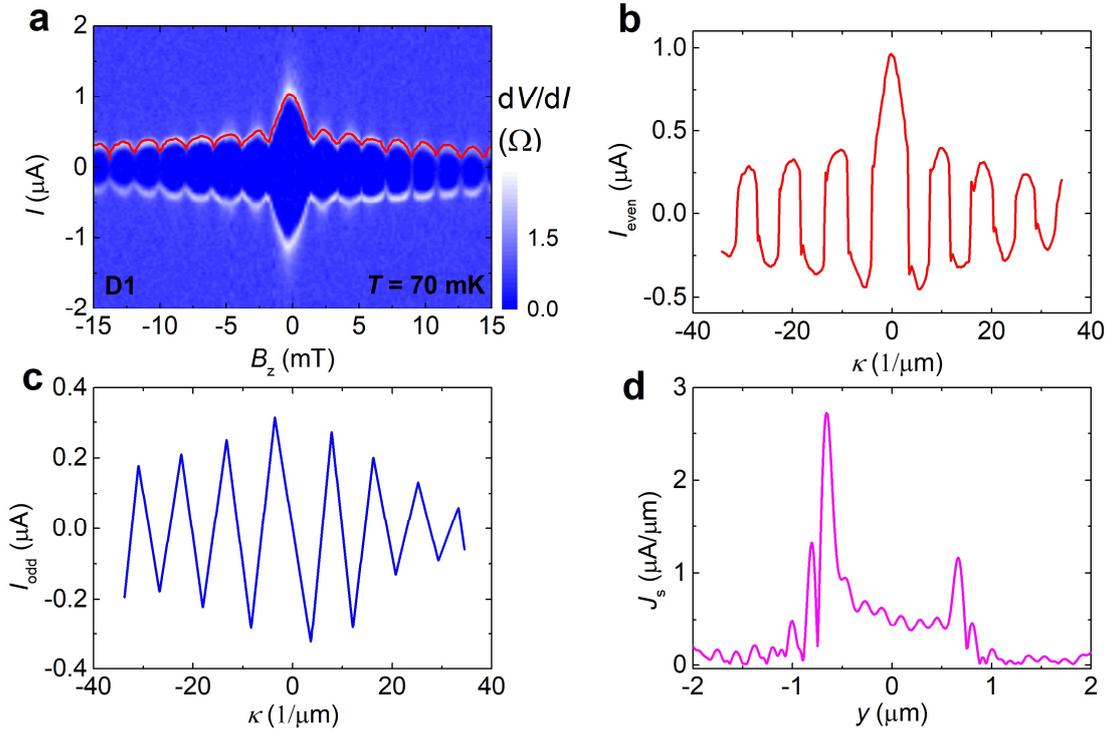

Fig. S1. **a**, SIP for D1, the same as Fig. 2a in the main text. The red curve illustrates $I_c(B_z)$ taken following the whitish envelope for an example. **b**, The even part of $I_c(\kappa)$ recovered from the red curve in **a**. **c**, The odd part of $I_c(\kappa)$ recovered from the red curve in **a**. **d**, Supercurrent density profile for $I_c(B_z)$ in **a**.

**Section II. Magnitude of $B_x$ to kill the bulk supercurrent for different devices.**

We note that the $B_x$ for killing the bulk supercurrent on D1 is much larger than D2, D3 and D4. We think it is caused by the detailed fabrication process (e.g., the quality of the superconductor films), because D2, D3 and D4 are fabricated simultaneously. However, D1 was fabricated and measured around six months earlier than them. In Fig. S2, we present another device D5, which was fabricated together with D1, and it also shows a larger $B_x$ for killing the bulk supercurrent.

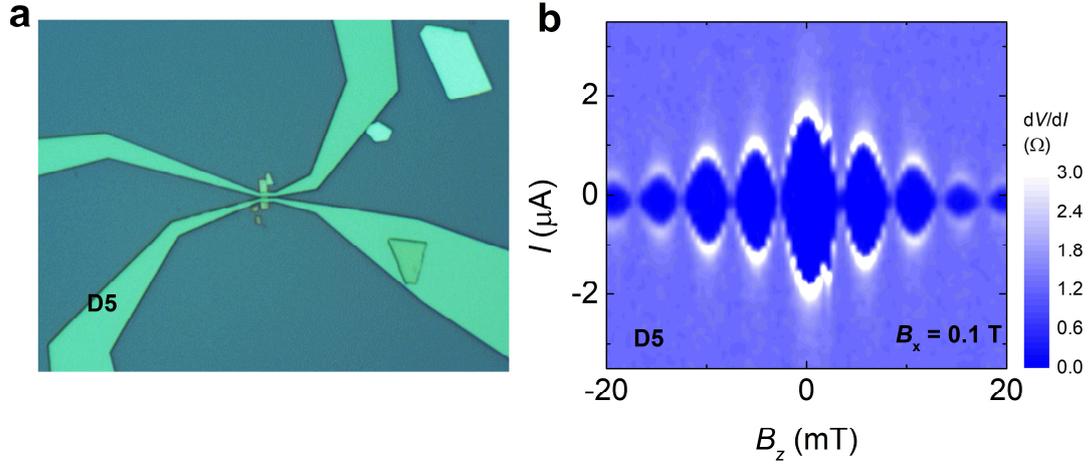

Fig. S2. **a**, The optical image for D5. **b**, SIP for D5 under $B_x = 0.1$ T.

**Section III. Anisotropy of bulk supercurrent between $B_x$ and $B_y$.**

As shown by Fig. 4 in the main text, the critical field of bulk supercurrent along $B_x$ is lower than $B_y$. We think it is a common phenomenon that has been reported on other Josephson junctions with a similar electrode configuration[1]. We attribute it to the vimineous shape of the electrodes that exhibit anisotropic demagnetization $N$[2]. It is natural to assume the proximity-induced superconducting region beneath the electrodes to be a stripe shape. The effective demagnetization can be approximated as[3]:

$$N^{-1} = 1 + \frac{3}{4}\frac{l_\parallel}{l_\perp}$$

where the symbol $l_\parallel$ represents the length along the magnetic field direction, $l_\perp$ is the length perpendicular to the magnetic field. Suppose $l_\parallel$ and $l_\perp$ are comparable to the size of the electrodes on the sample (2.1 μm × 0.5 μm), and $N$ is estimated to be around 0.84 for $B_x$ and 0.24 for $B_y$, which produce a large anisotropy.

**Section IV. Analysis of the critical side-surface supercurrent (side-surface-$I_c$) under $B_y$.**

As we mentioned in the main text, if the boundary supercurrent originates from the side surfaces (rather than the hinges), we can calculate its value at a certain $B_y$ through the Fraunhofer-like decay curve. To do so, we need to supress the contribution of the bulk supercurrent, which can be achieved by applying a small $B_z$ to render the Fraunhofer-like $1/|B_z|$ decay of the bulk supercurrent itself. Therefore, we inspect the side lobes of SIP in Fig. 2a in the main text, where the bulk supercurrent has been supressed due to the Fraunhofer diffraction in $B_z$. Of course, the larger the serial number $|n|$ of the side lobes, the heavier the suppression of the bulk supercurrent. Nevertheless, we assign the height of the side lobes (excluding the central lobe) of the SIP at $B_y = 0$ T in Fig. 2a as the critical supercurrent of the side-surface, side-surface-$I_c$. In the main text, we assume that the effective junction length of the side surface is comparable to the separation of the electrodes, i.e., ~300 nm. In fact, it could be underestimated if considering the flux focusing effect of the electrodes. We thus test several conditions here, as shown in Fig. S3, and side-surface-$I_c$ is always smaller than the height of the first lobe ~$0.21 I_c^0$, where $I_c^0$ is the side-surface-$I_c$ at $B_y = 0$ T. However, the experimental boundary-$I_c$ at $B_y = 0.2$ T is much larger than $0.21 I_c^0$ as shown in Fig. 5c in the main text, in contrast to the side-surface scenario.

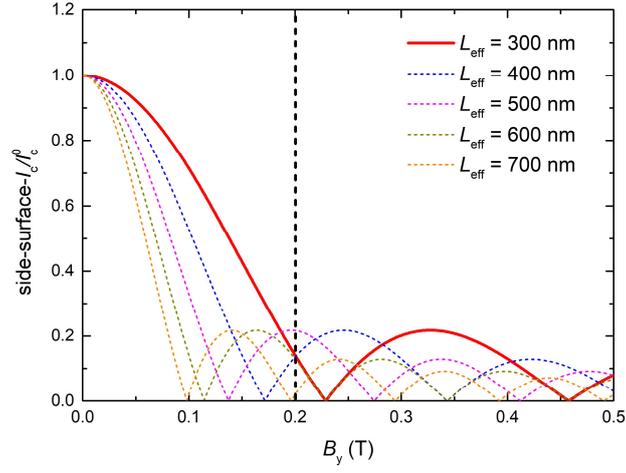

Fig. S3. Simulation of Fraunhofer-like curves for different effective junction lengths assuming a side-surface supercurrent.

**Section V. SIP for D1.**

Figure S4 shows the SIP for D1 at $B_y = 0.2$ T, which was used to extract the boundary-$I_c$ at $B_y = 0.2$ T for the side lobes presented in Fig. 5c in the main text.

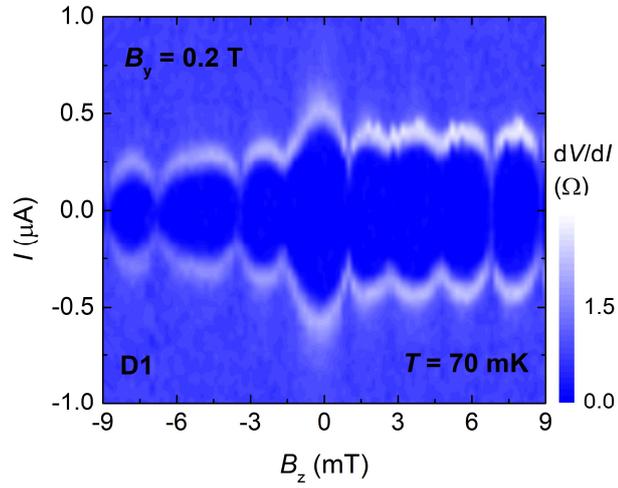

Fig. S4. SIP for D1 at $B_y = 0.2$ T.

**Section VI. Theoretical calculation.**

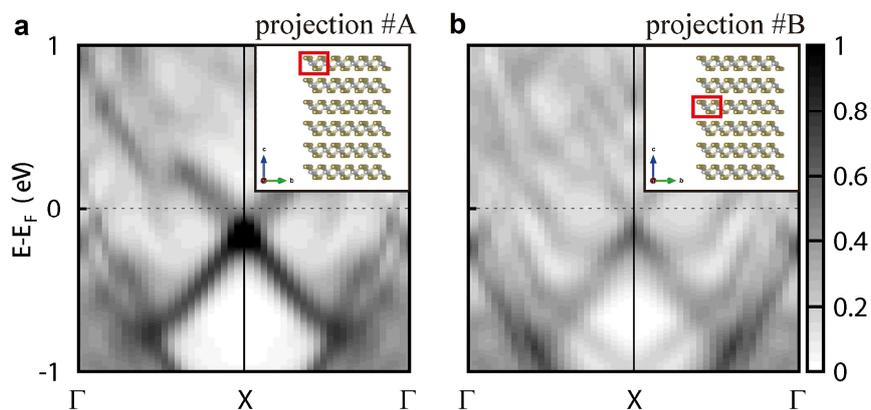

Fig. S5. The bands of **a**, hinge state and **b**, side-surface state. Compared with the side-surface atoms' projection (projection #B), the hinge atoms' projection (projection #A) contributes more around $E_F$, which indicates that the band contribution near the Fermi level is dominated by the hinge state rather than side-surface state.